\documentclass{article}

\usepackage{PRIMEarxiv}

\usepackage[utf8]{inputenc} 
\usepackage[T1]{fontenc}    
\usepackage{hyperref}       
\usepackage{url}            
\usepackage{booktabs}       
\usepackage{amsfonts}       
\usepackage{nicefrac}       
\usepackage{microtype}      
\usepackage{lipsum}
\usepackage{amsmath}
\usepackage{fancyhdr}       
\usepackage{graphicx}       
\graphicspath{{media/}}     

\title{NAWOA-XGBoost: A Novel Model for Early Prediction of Academic Potential in Computer Science Students
\thanks{\textit{\underline{Citation}}: 
\textbf{Authors. Title. Pages.... DOI:000000/11111.}} 
}

\author{
  Junhao Wei, Yanzhao Gu, Ran Zhang, Mingjing Huang, Jinhong Song, Yanxiao Li, Wenxuan Zhu, Yapeng Wang \\
  Faculty of Applied Sciences \\
  Macao Polytechnic University \\
  Macao, China\\
  \texttt{\{p2312195,p2311998,p2512396,P2314640,p2315937,P2525981,p2525620,yapengwang\}@mpu.edu.mo}  
   \AND
   Zikun Li  \\
   School of Economics and Management \\
   South China Normal University \\
   \texttt{18520610821@163.com} \\
   \And
   Zhiwen Wang \\
   School of Nursing \\
   Peking University \\
   \texttt{wzwjing@sina.com} \\
   \And
   Xu Yang*, Ngai Cheong* \\
   Faculty of Applied Sciences \\
   Macao Polytechnic University \\
   \texttt{\{xuyang,ncheong\}@mpu.edu.mo} \\
}

\begin{document}
\maketitle

\begin{abstract}
Whale Optimization Algorithm (WOA) suffers from limited global search ability, slow convergence, and tendency to fall into local optima, restricting its effectiveness in hyperparameter optimization for machine learning models. To address these issues, this study proposes a Nonlinear Adaptive Whale Optimization Algorithm (NAWOA), which integrates strategies such as Good Nodes Set initialization, Leader-Followers Foraging, Dynamic Encircling Prey, Triangular Hunting, and a nonlinear convergence factor to enhance exploration, exploitation, and convergence stability. Experiments on 23 benchmark functions demonstrate NAWOA's superior optimization capability and robustness. Based on this optimizer, an NAWOA-XGBoost model was developed to predict academic potential using data from 495 Computer Science undergraduates at Macao Polytechnic University (2009-2019). Results show that NAWOA-XGBoost outperforms traditional XGBoost and WOA-XGBoost across key metrics, including Accuracy (0.8148), Macro F1 (0.8101), AUC (0.8932), and G-Mean (0.8172), demonstrating strong adaptability on multi-class imbalanced datasets.
\end{abstract}


\section{Introduction}
With the rapid development of Educational Data Mining (EDM) \cite{bib1}, predicting students' future academic potential based on their early learning performance has become an important research direction for quality assurance, resource allocation, and personalized instruction in higher education. In the field of Computer Science (CS), students' academic records and behavioral data during their first and second years often provide strong indications of their foundational knowledge, learning capability, and future development potential. Therefore, establishing a reliable early academic potential prediction model is of significant importance.\par
In recent years, machine learning methods have been extensively applied to educational prediction tasks. Among these methods, XGBoost has been regarded as a suitable model due to its strong generalization ability and superior performance on structured data. However, the performance of XGBoost highly depends on its hyperparameter settings. Manual tuning is inefficient and prone to local optima, which restricts further improvements in predictive performance. To address this limitation, researchers started to pay attention on metaheuristic algorithms. Metaheuristic   algorithms, which are famous of their strong optimization abilities, have been introduced to path planning \cite{ipso}\cite{drrt}\cite{ahrrt}, engineering design, antenna design \cite{mrbmo}, and optimizing the hyperparameters of machine learning models. Among them, the Whale Optimization Algorithm (WOA) has attracted attention due to its simplicity, efficiency, and global search capability. Nevertheless, the conventional WOA tends to suffer from premature convergence and loss of population diversity in later iterations, limiting its ability to effectively explore complex parameter spaces.\par
Previous studies have integrated the classical WOA with XGBoost and achieved promising prediction results \cite{xgbwoa}. However, there remains considerable room for improvement. Building upon this foundation, the present study proposes a Nonlinear Adaptive Whale Optimization Algorithm (NAWOA), which incorporates strategies such as Good Nodes Set initialization, Leader-Followers Foraging, and Dynamic Encircling Prey to enhance global exploration ability and optimization stability. Experiments conducted on 23 benchmark functions validate the strong optimization capability of the proposed algorithm \cite{rwoa}\cite{gwoa}. Based on this enhanced optimizer, we construct an NAWOA-XGBoost academic potential prediction model and evaluate it using a real dataset of 495 undergraduate students majoring in Computer Science at Macao Polytechnic University (MPU) from 2009 to 2019. The dataset was preprocessed and provided by the institution, enabling direct use for training and comparative analysis.\par
Experimental results indicate that NAWOA can more effectively optimize the hyperparameters of XGBoost, thereby enhancing model stability and predictive performance in multi-class educational prediction tasks. The proposed model provides a valuable reference for early academic potential identification and personalized educational support in higher education institutions.

\section{WOA}
The Whale Optimization Algorithm (WOA), proposed by Mirjalili et al. in 2016, is a metaheuristic optimization technique inspired by the foraging behavior of humpback whales \cite{woa}. The algorithm models the whales' cooperative hunting strategies observed in nature, particularly their distinctive bubble-net feeding mechanism. WOA simulates three core behavioral patterns of humpback whales: search-for-prey, spiral-updating during bubble-net feeding, and encircling the prey. Although WOA is structurally simple, easy to implement, and has demonstrated good performance in various optimization applications, it still exhibits several limitations. First, WOA tends to suffer from a decline in population diversity during the later stages of iteration, which weakens its global search capability. Second, the algorithm often shows slow convergence, limited optimization accuracy in complex or high-dimensional problems, and a tendency to become trapped in local optima. Moreover, achieving a proper balance between global exploration and local exploitation highly depends on parameter settings, making WOA sensitive to parameter variations and reducing its stability in real-world applications.\par
To address these shortcomings, this study proposes an enhanced multi-strategy Whale Optimization Algorithm (NAWOA). The improved algorithm strengthens the balance between exploration and exploitation, increases search efficiency, and enhances optimization accuracy by integrating several newly designed strategies tailored to overcome the weaknesses of the original WOA.

\section{The Proposed NAWOA}
\subsection{Good Nodes Set initialization}
\begin{figure}[htbp]
    \centering
    \includegraphics[width=\textwidth]{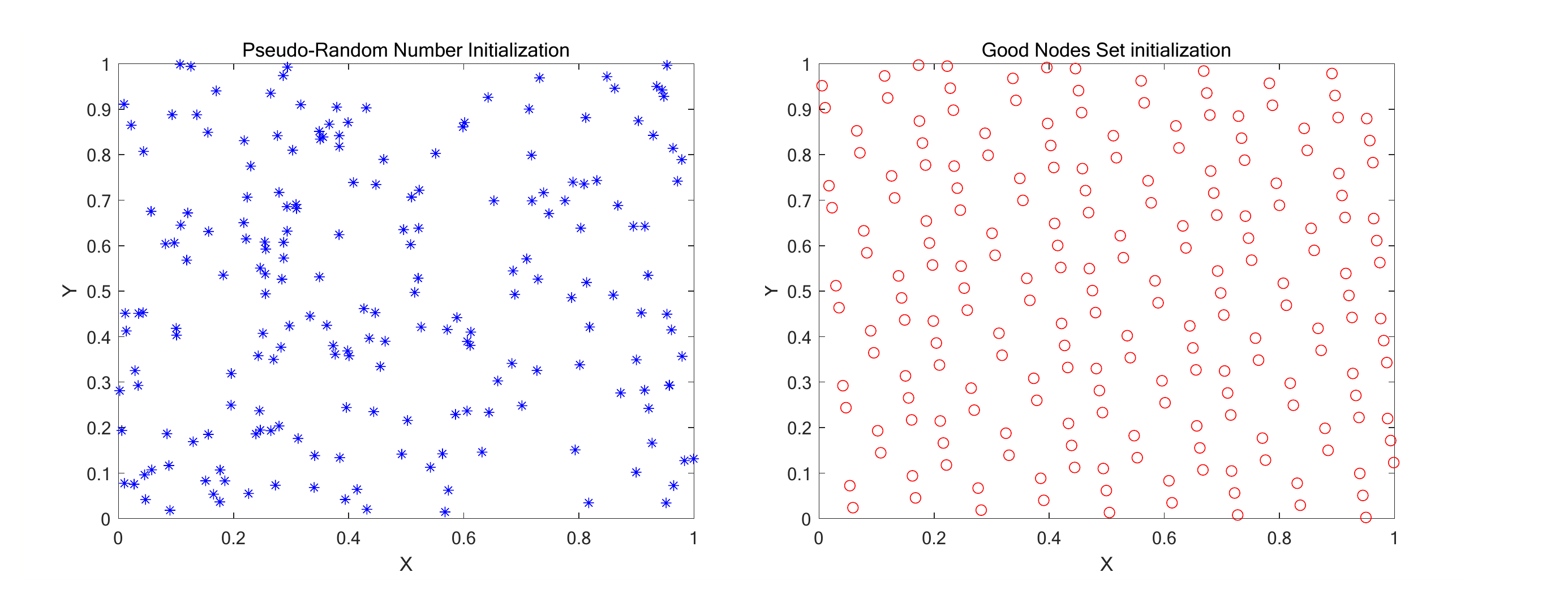}
    \caption{Pseodo-Random Number initialization vs Good Nodes Set
    initialization ($N$=200)} 
    \label{ini}
\end{figure}

In the classical WOA, the initial population is typically generated using pseudo-random numbers, as illustrated on the left side of Figure~\ref{ini}. Although this method is easy to implement and can quickly produce an initial set of solutions, the resulting individuals are often unevenly distributed, leading to low population diversity and noticeable clustering. \par
To overcome these limitations, NAWOA adopts the Good Nodes Set method for population initialization \cite{glnwoa}. The concept of the Good Nodes Set was introduced by Hua Luogeng, aiming to generate point sets with more uniform spatial distribution. A key advantage of this method is that its construction is dimension-independent—meaning that it maintains uniformity not only in two-dimensional spaces but also in high-dimensional environments. As shown on the right side of Figure~\ref{ini}, when the population size is set to $N$=200, the population generated by the Good Nodes Set exhibits a more homogeneous distribution compared to pseudo-random number initialization. This effectively avoids population clustering, increases coverage of the search space, and strengthens global exploration performance during early iterations.\par
Let $U^D$ denote the unit hypercube in a $D$-dimensional Euclidean space. The Good Nodes Set can be defined as:
\begin{equation}
    P_r^M=\{p(k)=(\{kr\},\{kr^2\},...,\{kr^D\})|k=1,2,...,M\}
    \label{eq1}
\end{equation}
where $\{x\}$ denotes the fractional part of $x$, $M$ is the number of nodes, $r>0$ is an offset parameter, $C(r,\varepsilon)$ is a constant depending on $r$, and $\varepsilon>0$ is a given constant.\par
When mapping the Good Nodes Set to the actual search space, suppose the lower and upper bounds of the $i^{th}$ dimension are $x_i^{min}$ and $x_i^{max}$, respectively. The mapping is performed as follows:
\begin{equation}
   x^i_{k}=x^i_{min}+p_i(k) \cdot (x^i_{max}-x^i_{min})
    \label{eq2}
\end{equation}
Through this mapping, the Good Nodes Set enables the generation of more uniformly distributed initial solutions within the search space, effectively enhancing the global exploration ability and search stability of NAWOA.

\subsection{Leader-Followers Foraging}
In the traditional WOA, the 'search for prey' process relies on randomly selecting an individual to update positions. Although this strategy increases stochasticity and diversity to some extent, it also introduces several limitations: the search directions of individuals lack consistency, convergence becomes unstable in later iterations, and the algorithm depends excessively on random factors, resulting in insufficient utilization of population information. Particularly in the late stage of optimization, individual updates often exhibit irregular oscillations, making the algorithm prone to premature convergence or entrapment in local optima.\par
To address these issues, this study draws inspiration from the natural hunting behavior of humpback whales, in which a leader guides the group toward the prey \cite{lsewoa}. Accordingly, we propose the Leader–Followers Foraging strategy. This strategy introduces a Leader (the current global best individual) as the primary driving force and incorporates the population mean position to guide individuals toward more structured and direction-aware searches, thereby improving the stability and effectiveness of the optimization process.\par
The position update model of the Leader–Followers Foraging strategy is defined as follows:
\begin{equation}
    {X}(t+1)=(1-\frac tT) \cdot {X}^*(t)+|{X}_R(t)-{X}^*(t)|
     \label{eq3}
\end{equation}
where $t$ is the current iteration number; $T$ is the maximum number of iterations; ${X}_i$ is the position of the whale individual; ${X}^*$ denotes the position of the current best solution.\par
${X}_R$ is the average position of all whale individuals:
\begin{equation}
    {X}_R(t)=\frac1N\sum_{i=1}^N{X}_i(t)
     \label{eq4}
\end{equation}
where $N$ is the population size.\par
Under this update mechanism, each whale individual is jointly influenced by the Leader’s position and the overall population distribution. The Leader provides a clear global convergence direction, while the population mean position constrains and regulates the search range, ensuring that the update process maintains both directionality and stability. This strategy effectively reduces instability caused by excessive randomness in the search process, enabling individuals to rely more on population structural information and preventing overly weak or overly strong mutual influence among individuals. As a result, the overall coordination of the population search is significantly enhanced. The Leader-Followers Foraging strategy strengthens global exploration while improving convergence stability, enabling a more precise and efficient optimization process.

\subsection{Dynamic Encircling Prey}
In the original WOA, during the encircling prey phase, individual positions are updated mainly based on the Euclidean distance between the current best solution and each whale. Although this linear shrinking mechanism is simple, its search trajectories are relatively monotonic and lack flexibility. In the later stages of iteration, the movement amplitude becomes too small, making individuals prone to being trapped in local optima. Moreover, when the contraction coefficient $A$ becomes small, the update range of individuals is significantly restricted, further reducing the algorithm’s ability to escape local regions and negatively affecting overall convergence performance.\par
To enhance the diversity and dynamic behavior of the encircling phase, this study introduces a spiral flight mechanism and proposes the Dynamic Encircling Prey strategy \cite{tswoa} \cite{lswoa}. This strategy incorporates a spiral flight step into the traditional encircling mechanism, enabling whale individuals to follow diverse and nonlinear trajectories when approaching the prey. This improves both the exploration capability and the ability to escape from local optima. The modeling of the Dynamic Encircling Prey strategy is as follows.\par
First, the distance between the whale individual and the Leader is calculated:
 \begin{equation}
	{D}=|{C}\cdot {X}^*(t)-{X}(t)|
	 \label{eq5}
 \end{equation}
where ${A}$ and ${C}$ are coefficient vectors.\par 
Then, a spiral-flight-based disturbance term is constructed using random variables:
 \begin{equation}
	L=2\cdot r-1
	 \label{eq6}
 \end{equation}
 \begin{equation}
	Z=e^{k\cdot cos(\pi\cdot(1-\frac tT))}
	 \label{eq7}
 \end{equation}
where $Z$ represents the Spiral flight step size; $s$ and $k$ are spiral coefficients; and $r$ is a random number between 0 and 1.\par
Finally, the position update model is formulated as:
\begin{equation}
	{X}(t+1)={X}^*(t)+e^{Z\cdot L}\cdot cos(2\pi L)\cdot|{A}\cdot {D}|
	 \label{eq5}
 \end{equation}
In this strategy, the spiral step size $Z$ dynamically changes with the iterations, allowing whale individuals to exhibit different disturbance magnitudes at different optimization stages. The random variable L introduces directional diversity, generating rich encircling trajectories. Through this nonlinear flight mechanism, individuals can not only perform thorough local exploration when approaching the prey but also gain a higher probability of escaping local optima, thus avoiding premature convergence. The Dynamic Encircling Prey strategy significantly enhances flexibility during the encircling phase, equipping the algorithm with stronger dynamic exploration ability when approaching the optimal solution, thereby contributing to improved convergence accuracy and robustness.

\subsection{Triangular Hunting}
In the original WOA, the spiral updating mechanism adjusts individual positions around the current best solution, guiding whales to approach the prey along a fixed spiral trajectory. Although this strategy mimics the bubble-net spiral feeding behavior of humpback whales, it becomes overly dependent on the current optimal solution in the later stages of iteration. The disturbance range becomes limited, restricting the ability of individuals to perform more complex spatial exploration. As the algorithm approaches the optimal region, the use of a single spiral path increases the risk of falling into local optima, limiting both fine-grained search capability and the ability to escape suboptimal regions. To enhance the spatial exploration ability during the spiral hunting phase, this study proposes the Triangular Hunting strategy. By introducing dynamically scaled random disturbances, multi-directional angular deviations, and a triangular step length, whale individuals can generate richer and more complex movement trajectories when approaching the prey. This significantly improves both local search precision and the ability to escape local optima. The modeling of the Triangular Hunting strategy is as follows.\par
First, a dynamic scaling factor $r_1$ and a random scaling factor $R$ are constructed:
\begin{equation}
    r_1=0.1\cdot(1-\frac tT)
     \label{eq6}
\end{equation}
\begin{equation}
    r=rand\cdot r_1
     \label{eq7}
\end{equation}
where $Rand$ denotes a random number between 0 and 1; $r_1$ gradually decreases as iterations progress, enabling wide-range exploration in early stages and more stable local exploitation in later stages.\par
The distance term consists of two parts: the straight-line distance to the optimal individual and a random step length:
\begin{equation}
    L_1=|X^*(t)-X(t)|
     \label{eq8}
\end{equation}
\begin{equation}
    L_2=D^{\prime}\cdot rand
     \label{eq9}
\end{equation}
A random angle $\gamma$ is introduced to generate asymmetric perturbations:
\begin{equation}
    \gamma=2\cdot\pi\cdot rand
     \label{eq10}
\end{equation}
Then the triangular step length $L$ is computed:
\begin{equation}
    L=\sqrt{|L_1\cdot L_1+L_2\cdot L_2-2\cdot L_1\cdot L_2\cdot cos(\gamma)|}
     \label{eq11}
\end{equation}
Finally, the position update model of Triangular Hunting is expressed as:
\begin{equation}
    X(t+1)=X^*(t)\cdot D^{\prime}+r\cdot L+e^{Z\cdot L}\cdot\cos(2\pi L)\cdot|A\cdot D|
     \label{eq31}
\end{equation}
In this strategy, the triangular step $L$ enables individuals to deviate from the traditional spiral path and explore in multiple directions, forming more flexible hunting trajectories. The dynamic scaling factor $r_1$ ensures a smooth transition from broad early exploration to refined late exploitation. The random angle $\gamma$ increases the asymmetry of perturbations, preventing individuals from collapsing into a single spiral path, thereby enhancing diversity and robustness. By combining triangular disturbances with spiral update behavior, the Triangular Hunting strategy enables the population to maintain a convergent trend while performing more comprehensive and flexible neighborhood exploration. As a result, the algorithm achieves significantly improved local optimization capability and a stronger ability to escape local optima.

\subsection{Nonlinear convergence factor $a$}
Building upon the aforementioned enhancement strategies, although the search capability of the algorithm has been significantly improved, the linear convergence factor $a$ used in the original WOA remains insufficient for meeting the regulatory requirements of the enhanced NAWOA. In the classical method, $a$ decreases linearly from 2 to 0, and this relatively smooth transition results in limited differentiation of convergence behavior across different stages of the iteration process. In particular, the slow decline in the later iterations may restrict the algorithm’s convergence efficiency and search precision.\par
To achieve more flexible control over the exploration–exploitation balance and better align with the newly introduced mechanisms of NAWOA, this study adopts a nonlinear convergence factor update strategy based on a Sigmoid curve. This update pattern enables $a$ to follow a 'slow–fast–slow' transition during the optimization process, which better matches the dynamic requirements of real-world optimization behavior. The nonlinear convergence factor is modeled as follows:
\begin{equation}
    a=2-\frac2{1+e^{-25(\frac tT-0.5)}}
    \label{eq36}
\end{equation}
This Sigmoid-based update mechanism allows $a$ to adjust dynamically according to the needs of different optimization phases, resulting in a more adaptive control of search intensity. It establishes a more reasonable balance among global exploration, convergence speed, and local exploitation, thereby enhancing the overall optimization performance and stability of the algorithm.

\section{Experiments}
\subsection{Comparative Experiment on Benchmark Functions}
\begin{figure*}[htbp]
    \centering
    \includegraphics[width=\textwidth]{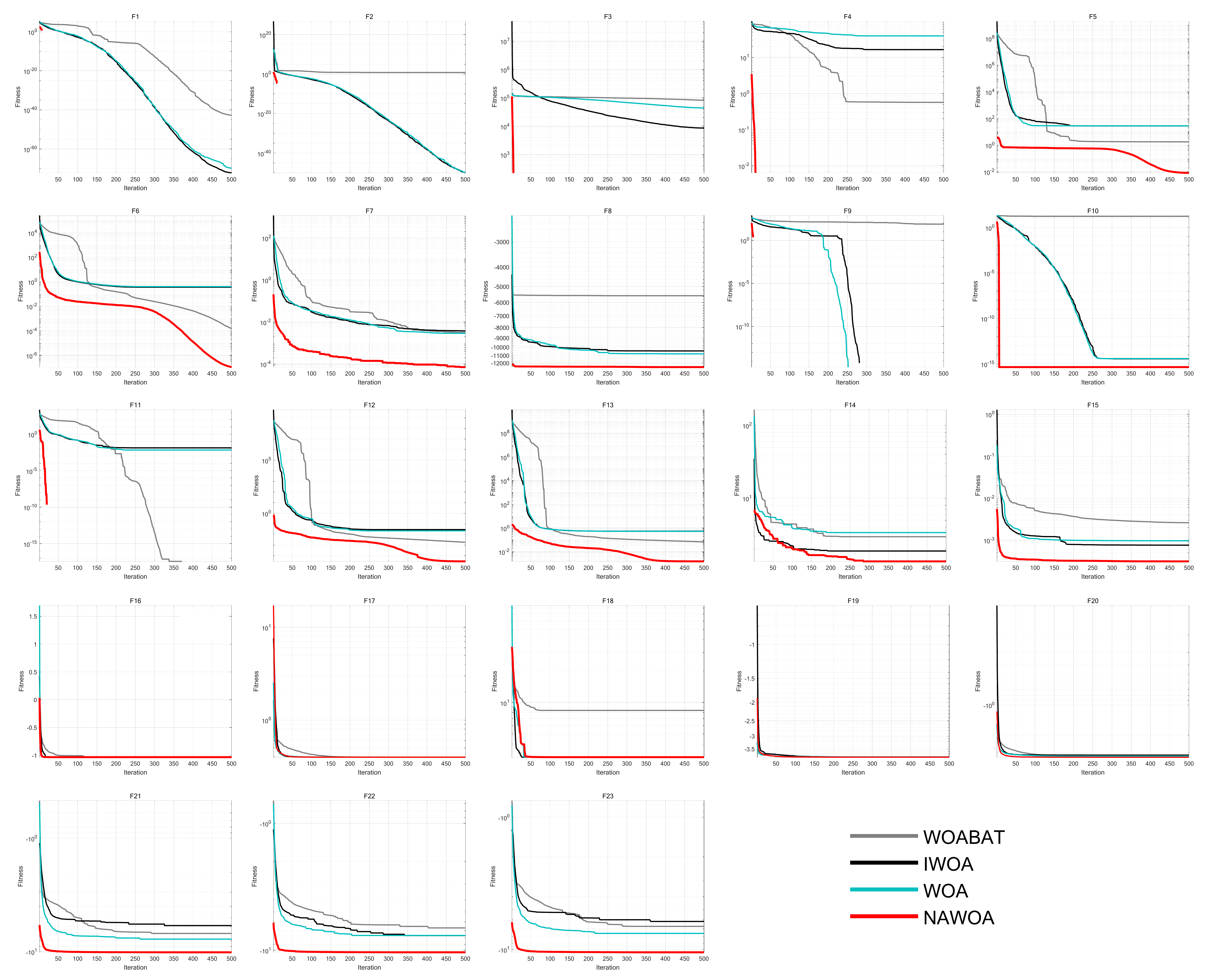}
    \caption{Iterative curves of the algorithms on Benchmark Functions.}
    \label{differ}
\end{figure*}

To evaluate the optimization capability of NAWOA, this study selected WOABAT \cite{WOABAT}, IWOA \cite{IWOA}, and WOA \cite{woa} for comparison with NAWOA on the 23 benchmark functions listed in Table~\ref{table1}. The parameter settings for each algorithm were shown in Table~\ref{table2}. The number of iterations was uniformly set to $T$=500, and the population size was set to $N$=30. Each algorithm was independently executed 30 times on the 23 benchmark functions, and the average fitness ($Ave$) and standard deviation ($Std$) were recorded for performance analysis. The experimental results are shown in Figure~\ref{differ} and Table~\ref{differ2}.The results indicate that NAWOA achieves the overall best performance among all algorithms. Compared to the original WOA, NAWOA significantly improves overall performance. As shown in Figure~\ref{differ} and Table~\ref{differ2}, NAWOA attains the best results in terms of both average fitness and standard deviation for the majority of benchmark functions, demonstrating its strong adaptability and robustness in solving various types of optimization problems.
\begin{table}[h]
    \centering
    \caption{Classical Benchmark Functions}
    \resizebox{\linewidth}{!}{
    \begin{tabular}{c c c c c}
    \hline
    \textbf{Function} & \textbf{Function's Name} & \textbf{Type} & \textbf{Dimension} & \textbf{Best Value} \\
    \hline
    F1 & Sphere & Uni-modal & 30 & 0 \\
    F2 & Schwefel's Problem 2.22 & Uni-modal & 30 & 0 \\
    F3 & Schwefel's Problem 1.2 & Uni-modal & 30 & 0 \\
    F4 & Schwefel's Problem 2.21 & Uni-modal & 30 & 0 \\
    F5 & Generalized Rosenbrock's Function & Uni-modal & 30 & 0 \\
    F6 & Step Function & Uni-modal & 30 & 0 \\
    F7 & Quartic Function & Uni-modal & 30 & 0 \\
    F8 & Generalized Schwefel's Function & Multi-modal & 30 & -12569.5 \\
    F9 & Generalized Rastrigin's Function & Multi-modal & 30 & 0 \\
    F10 & Ackley's Function & Multi-modal & 30 & 0 \\
    F11 & Generalized Griewank's Function & Multi-modal & 30 & 0 \\
    F12 & Generalized Penalized Function 1 & Multi-modal & 30 & 0 \\
    F13 & Generalized Penalized Function 2 & Multi-modal & 30 & 0 \\
    F14 & Shekel's Foxholes Function & Multi-modal & 2 & 0.998 \\
    F15 & Kowalik's Function & Multi-modal & 4 & 0.0003075 \\
    F16 & Six-Hump Camel-Back Function & Compositional & 2 & -1.0316 \\
    F17 & Branin Function & Compositional & 2 & 0.398 \\
    F18 & Goldstein-Price Function & Compositional & 2 & 3 \\
    F19 & Hartman's Function 1 & Compositional & 3 & -3.8628 \\
    F20 & Hartman's Function 2 & Compositional & 6 & -3.32 \\
    F21 & Shekel's Function 1 & Compositional & 4 & -10.1532 \\
    F22 & Shekel's Function 2 & Compositional & 4 & -10.4029 \\
    F23 & Shekel's Function 3 & Compositional & 4 & -10.5364 \\
    \hline
    \label{table1}
    \end{tabular}}
\end{table}

\begin{table}[h]
    \centering
    \caption{Parameter Settings for different metaheuristic algorithms}
    \resizebox{\linewidth}{!}{
    \begin{tabular}{c c c}
        \hline
        \textbf{Algorithm} & \textbf{Parameter} & \textbf{Value} \\
        \hline
        WOABAT & Convergence Factor $a$ & 2 decrease to 0 \\
        & Spiral Factor $b$ & 1 \\ 
        IWOA & Convergence Factor $a$ & 2 decrease to 0 \\
        & Spiral Factor $b$ & 1 \\ 
		WOA & Convergence Factor $a$ & 2 decrease to 0 \\
        & Spiral Factor $b$ & 1 \\ 
        NAWOA & Convergence Factor $a$ & 2 decrease to 0 \\
        & Spiral Factor $b$ & 1 \\
        & $k$ & 1 \\ 
        \hline
        \label{table2}
    \end{tabular}}
\end{table}

\begin{table*}[htbp]
\centering
\caption{Comparison Results of the Algorithms on Benchmark Functions.}
\resizebox{0.8\textwidth}{!}{
\begin{tabular}{cccccc}
\hline
Function & Metrics & WOABAT & IWOA & WOA & NAWOA \\
\hline
F1  & Ave & 1.5751E-43 & 5.3971E-73 & 1.4454E-70 & 0.0000E+00 \\
    & Std & 3.8858E-43 & 2.0814E-72 & 7.9093E-70 & 0.0000E+00 \\
F2  & Ave & 6.0000E+00 & 9.0122E-51 & 1.5064E-50 & 0.0000E+00 \\
    & Std & 1.4044E+01 & 3.2837E-50 & 7.8910E-50 & 0.0000E+00 \\
F3  & Ave & 8.2227E+04 & 8.5951E+03 & 4.3898E+04 & 0.0000E+00 \\
    & Std & 2.6152E+04 & 2.1756E+04 & 1.1864E+04 & 0.0000E+00 \\
F4  & Ave & 5.6694E-01 & 1.6411E+01 & 3.9289E+01 & 0.0000E+00 \\
    & Std & 2.8810E+00 & 1.7954E+01 & 2.8674E+01 & 0.0000E+00 \\
F5  & Ave & 1.7564E+00 & 2.7791E+01 & 2.7851E+01 & 8.3346E-03 \\
    & Std & 6.6783E+00 & 5.0361E-01 & 3.7287E-01 & 9.3774E-03 \\
F6  & Ave & 1.4016E-04 & 3.5344E-01 & 3.9484E-01 & 1.0698E-07 \\
    & Std & 4.9084E-04 & 1.7209E-01 & 1.9620E-01 & 1.4777E-07 \\
F7  & Ave & 3.1190E-03 & 3.7522E-03 & 2.8885E-03 & 7.1110E-05 \\
    & Std & 5.9318E-03 & 5.0444E-03 & 2.7763E-03 & 5.9531E-05 \\
F8  & Ave & -5.5492E+03 & -1.0442E+04 & -1.0796E+04 & -1.2569E+04 \\
    & Std & 1.3059E+02 & 1.6224E+03 & 1.6168E+03 & 4.0759E-03 \\
F9  & Ave & 7.6652E+01 & 0.0000E+00 & 0.0000E+00 & 0.0000E+00 \\
    & Std & 7.7475E+01 & 0.0000E+00 & 0.0000E+00 & 0.0000E+00 \\
F10 & Ave & 1.6637E+01 & 3.6415E-15 & 3.8784E-15 & 4.4409E-16 \\
    & Std & 7.5676E+00 & 2.6960E-15 & 2.5523E-15 & 0.0000E+00 \\
F11 & Ave & 0.0000E+00 & 1.3375E-02 & 6.7765E-03 & 0.0000E+00 \\
    & Std & 0.0000E+00 & 4.3147E-02 & 3.7117E-02 & 0.0000E+00 \\
F12 & Ave & 1.8335E-03 & 2.8986E-02 & 2.2238E-02 & 2.9326E-05 \\
    & Std & 2.9019E-03 & 2.6139E-02 & 1.6859E-02 & 5.9552E-05 \\
F13 & Ave & 6.7397E-02 & 5.3773E-01 & 5.2470E-01 & 1.5373E-03 \\
    & Std & 1.0271E-01 & 2.6867E-01 & 2.6388E-01 & 4.7569E-03 \\
F14 & Ave & 2.9638E+00 & 1.8863E+00 & 3.3813E+00 & 1.3616E+00 \\
    & Std & 3.0494E+00 & 1.8225E+00 & 3.8167E+00 & 8.4290E-01 \\
F15 & Ave & 2.6078E-03 & 7.6137E-04 & 9.7030E-04 & 3.1374E-04 \\
    & Std & 5.3040E-03 & 4.9658E-04 & 1.2744E-03 & 1.5148E-05 \\
F16 & Ave & -1.0316E+00 & -1.0316E+00 & -1.0316E+00 & -1.0316E+00 \\
    & Std & 1.6177E-12 & 2.2765E-09 & 1.1327E-09 & 2.1993E-15 \\
F17 & Ave & 3.9789E-01 & 3.9790E-01 & 3.9790E-01 & 3.9789E-01 \\
    & Std & 1.3440E-10 & 2.6228E-05 & 1.4866E-05 & 1.5265E-13 \\
F18 & Ave & 8.4000E+00 & 3.0000E+00 & 3.0001E+00 & 3.0000E+00 \\
    & Std & 1.0985E+01 & 5.2176E-05 & 1.5524E-04 & 2.6580E-06 \\
F19 & Ave & -3.8628E+00 & -3.8526E+00 & -3.8548E+00 & -3.8624E+00 \\
    & Std & 7.5408E-06 & 1.3463E-02 & 1.3839E-02 & 1.1498E-03 \\
F20 & Ave & -3.2795E+00 & -3.1688E+00 & -3.2425E+00 & -3.3220E+00 \\
    & Std & 6.1567E-02 & 6.4067E-02 & 1.0904E-01 & 1.3247E-08 \\
F21 & Ave & -6.9245E+00 & -5.8987E+00 & -7.7704E+00 & -1.0153E+01 \\
    & Std & 2.4987E+00 & 1.9191E+00 & 2.8096E+00 & 3.4786E-11 \\
F22 & Ave & -6.6823E+00 & -7.6745E+00 & -7.6676E+00 & -1.0403E+01 \\
    & Std & 2.4774E+00 & 2.6469E+00 & 3.2685E+00 & 3.2957E-11 \\
F23 & Ave & -6.7076E+00 & -6.1580E+00 & -7.5781E+00 & -1.0536E+01 \\
    & Std & 2.8404E+00 & 2.2182E+00 & 3.3006E+00 & 3.5188E-11 \\
\hline
\end{tabular}
}
\label{differ2}
\end{table*}

\subsection{Comparison experiment on metaheuristic algorithms combining XGBoost}
\subsubsection{Dataset Description}
This study uses the student dataset from the Computer Science program at Macao Polytechnic University (MPU) covering the years 2009–2019, consistent with datasets used in related literature. The dataset contains multi-dimensional information of 495 undergraduate students, including academic performance, learning behavior, and course outcomes, and classifies students into different academic potential levels. The data were preprocessed by the university and are ready for model training and validation. The dataset has the following characteristics:\par
\begin{itemize}
    \item Multi-class classification problem: Different academic potential levels constitute a three-class prediction task;
    \item Diverse feature dimensions: Includes structured features such as course grades and learning trajectories;
    \item Real educational scenario data: Sourced from authentic teaching environments, providing representativeness and research value.
\end{itemize}
To ensure fairness, all models were trained and evaluated under the same training-test split conditions.

\subsubsection{Comparative Methods}
The three methods compared in the experiments are:\par
\begin{itemize}
    \item XGBoost: Baseline model without hyperparameter optimization \cite{xgb};
    \item WOA-XGBoost: Hyperparameters optimized using the traditional WOA;
    \item NAWOA-XGBoost: Hyperparameters optimized using the proposed NAWOA.
\end{itemize}
All models were evaluated using Accuracy, Macro Precision, Macro Recall, Macro F1, AUC, and G-Mean to comprehensively assess their adaptability to multi-class educational prediction tasks.

\subsubsection{Experimental Results and Analysis}
\begin{figure}[htbp]
    \centering
    \includegraphics[width=0.8\textwidth]{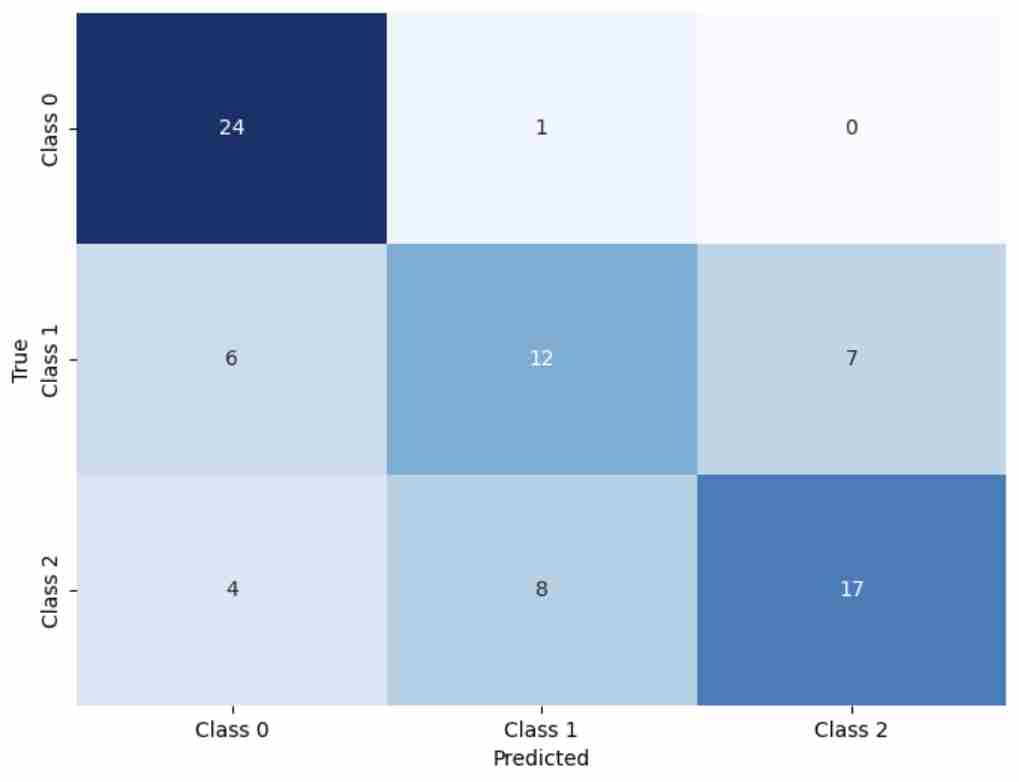}
    \caption{Confusion matrix of original XGBoost.}
    \label{XGB_matrix}
\end{figure}

\begin{figure}[htbp]
    \centering
    \includegraphics[width=0.8\textwidth]{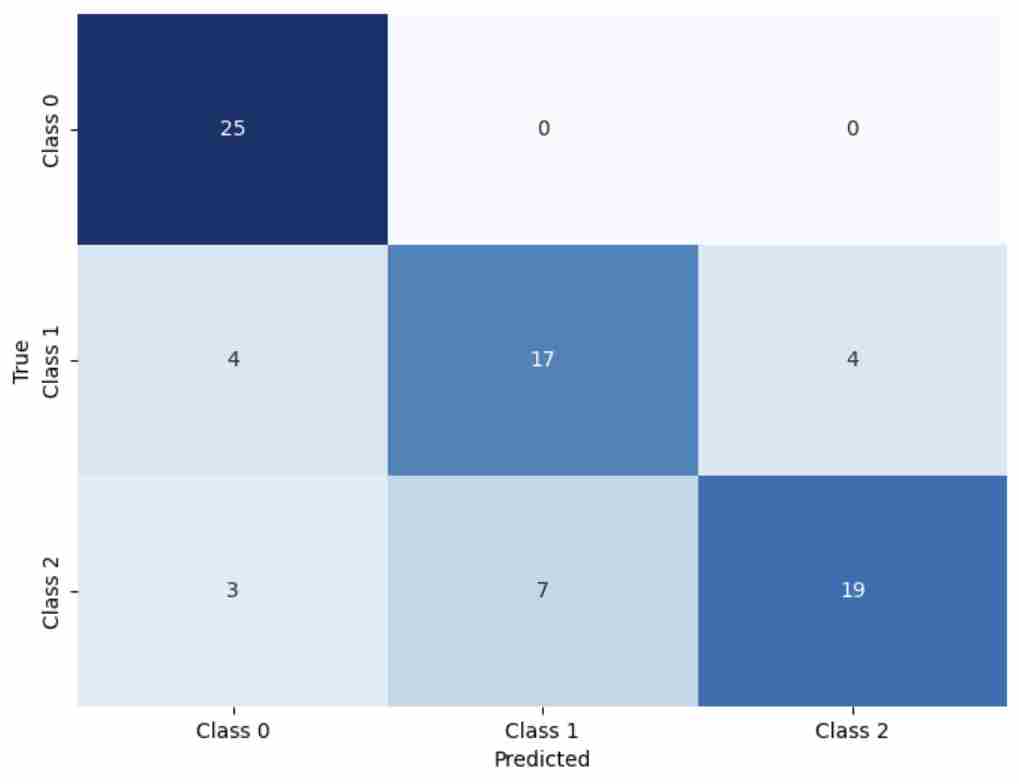}
    \caption{Confusion matrix of WOA-XGBoost.}
    \label{WOA_matrix}
\end{figure}

\begin{figure}[htbp]
    \centering
    \includegraphics[width=0.8\textwidth]{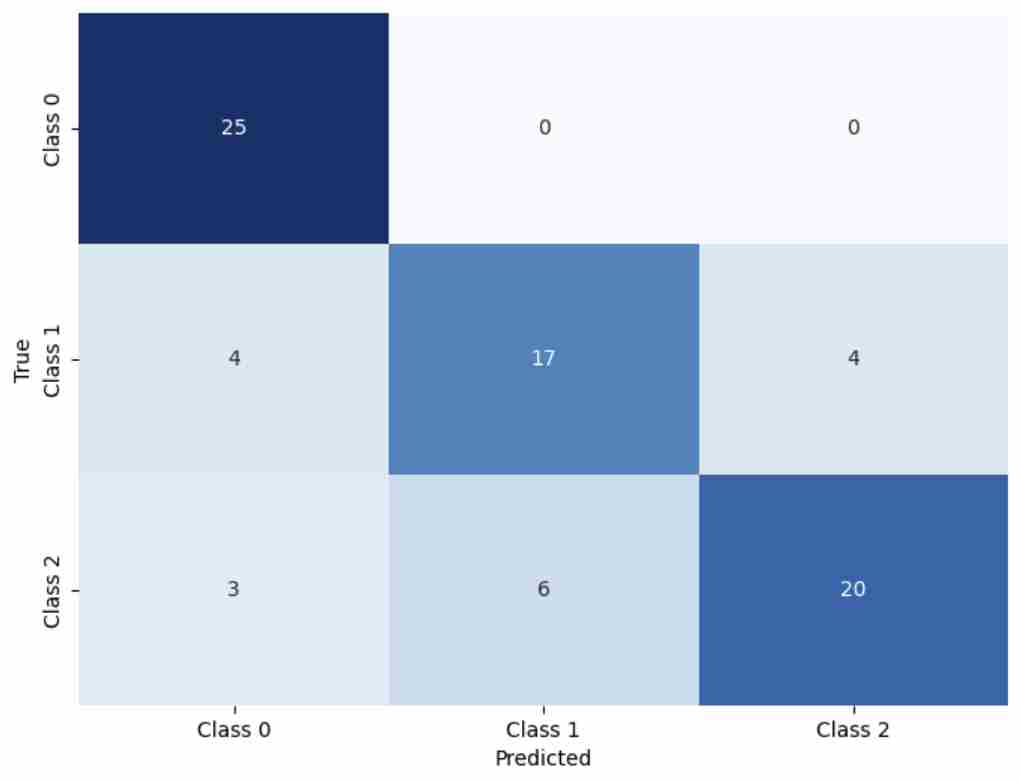}
    \caption{Confusion matrix of NAWOA-XGBoost.}
    \label{NAWOA_matrix}
\end{figure}

Table~\ref{xgb_compare_metrics}, Figure~\ref{XGB_matrix}, Figure~\ref{WOA_matrix} and Figure~\ref{NAWOA_matrix} present the experimental results of the three methods. It can be observed that NAWOA-XGBoost achieves the best performance across all evaluation metrics, outperforming both XGBoost and WOA-XGBoost.
\begin{table}[ht]
\centering
\caption{Comparison of Different methods}
\begin{tabular}{llll}
\hline
\textbf{Metrics} & \textbf{XGBoost} & \textbf{WOA-XGBoost} & \textbf{NAWOA-XGBoost} \\
\hline
Accuracy          & 6.7089E-01 & 7.6215E-01 & \textbf{8.1481E-01} \\
Macro Precision   & 6.6188E-01 & 7.6189E-01 & \textbf{8.1457E-01}\\
Macro Recall      & 6.7540E-01 & 7.6839E-01 & \textbf{8.1989E-01} \\
Macro F1 Score    & 6.5894E-01 & 7.5728E-01 & \textbf{8.1008E-01} \\
Micro F1 Score    & 6.7089E-01 & 7.6215E-01 & \textbf{8.1481E-01} \\
ROC AUC Score     & 8.5480E-01 & 8.5740E-01 & \textbf{8.9324E-01} \\
Gmean             & 6.6861E-01 & 7.6513E-01 & \textbf{8.1722E-01} \\
\hline
\end{tabular}
\label{xgb_compare_metrics}
\end{table}

\begin{itemize}
    \item Significant improvement in overall classification performance: The Accuracy of NAWOA-XGBoost reaches 0.8148, representing an improvement of approximately 21.5\% over the original XGBoost (0.6709) and further improvement over WOA-XGBoost (0.7621), indicating the enhanced global search capability of the improved algorithm;
    \item Comprehensive improvement in macro-average metrics: Macro Precision, Macro Recall, and Macro F1 reach 0.8146, 0.8199, and 0.8101, respectively, which are key indicators for evaluating imbalanced multi-class tasks. Compared to WOA-XGBoost, these metrics improve by approximately 5.3\%, 4.7\%, and 5.2\%, showing that NAWOA achieves more stable classification performance across multiple classes;
    \item Noticeable improvement in AUC: The AUC increases from 0.8548 (XGBoost) to 0.8932 (NAWOA-XGBoost), an improvement of nearly 4.5\%, indicating stronger overall discriminative ability;
    \item Significant improvement in G-Mean, achieving more balanced classification: G-Mean rises from 0.6686 (XGBoost) to 0.8172 (NAWOA-XGBoost), an increase of approximately 22\%, demonstrating more balanced performance across all classes.
\end{itemize}
These results indicate that strategies introduced in NAWOA, including Good Nodes Set initialization, Leader-Followers Foraging, Dynamic Encircling Prey, Triangular Hunting, and a nonlinear convergence factor, significantly enhance the algorithm’s exploration capability, ability to escape local optima, and convergence stability. Consequently, NAWOA-XGBoost demonstrates superior performance in predicting academic potential, accurately distinguishing students with different levels of potential. The experimental results fully validate the effectiveness of NAWOA in optimizing XGBoost, confirming the practical value and superior performance of the proposed method.

\section{Conclusion}
This study addresses the limitations of the traditional Whale Optimization Algorithm (WOA) in terms of search capability, convergence speed, and ability to escape local optima by proposing an improved algorithm, NAWOA. The algorithm incorporates multiple strategies, including Good Nodes Set initialization, Leader-Followers Foraging, Dynamic Encircling Prey, Triangular Hunting, and a nonlinear convergence factor, effectively enhancing global exploration, local exploitation, and overall convergence stability. Comparative experiments on the majority of benchmark functions demonstrate that NAWOA achieves superior performance in both average fitness and standard deviation, exhibiting stronger optimization capability and robustness.\par
To further validate the practical applicability of NAWOA, it was integrated with XGBoost for academic potential prediction and compared with traditional XGBoost and WOA-XGBoost. Experimental results based on real student data from the Computer Science program at Macau University of Science and Technology (2009–2019) indicate that NAWOA-XGBoost achieves the best performance across multiple key metrics, including Accuracy, Macro Precision, Macro Recall, Macro F1, AUC, and G-Mean. Specifically, Accuracy improved to 0.8148, Macro F1 to 0.8101, AUC to 0.8932, and G-Mean significantly increased to 0.8172, validating NAWOA’s strong adaptability and classification capability on multi-class imbalanced datasets. Overall, the results demonstrate that the proposed NAWOA not only significantly enhances hyperparameter optimization but also improves the stability and generalization ability of predictive models, underscoring its research significance and practical value.\par
Despite NAWOA’s promising performance, several directions warrant further investigation. First, this study primarily validated the algorithm on continuous optimization problems and structured educational data; future work could extend NAWOA to more complex problem types, such as combinatorial optimization, multi-objective optimization, and dynamic optimization environments, to further assess its generality. Second, the strategies in NAWOA are currently based on fixed mechanism designs; future research could explore adaptive parameter tuning or deep learning-assisted dynamic strategy generation to enhance the algorithm’s adaptability across diverse task environments. In addition, the current study relies on a single dataset; extending the validation to multi-institutional, multi-major, or cross-regional educational datasets would improve the model’s universality and robustness. Finally, NAWOA-XGBoost could be further combined with graph neural networks, temporal models, or more advanced ensemble methods to explore additional potential applications in educational assessment, student behavior analysis, and learning path prediction.

\section{Acknowledgment}
The supports provided by Macao Polytechnic University (MPU Grant no: RP/FCA-03/2022; RP/FCA-06/2022) and Macao Science and Technology Development Fund (FDCT Grant no: 0044/2023/ITP2) enabled us to conduct data collection, analysis, and interpretation, as well as cover expenses related to research materials and participant recruitment. MPU and FDCT investment in our work have significantly contributed to the quality and impact of our research findings.

\bibliographystyle{unsrt}  
\bibliography{references}

@article{woa,
  title={The whale optimization algorithm},
  author={Mirjalili, Seyedali and Lewis, Andrew},
  journal={Advances in engineering software},
  volume={95},
  pages={51--67},
  year={2016},
  publisher={Elsevier}
}

@article{WOABAT,
  title={Hybrid WOA-BAT Optimization for Performance Enhancement of Coexisting WBANs},
  author={Nandagopal, Priyanka},
  year={2021}
}

@inproceedings{ipso,
  title={Adaptive position updating particle swarm optimization for UAV path planning},
  author={Wei, Junhao and Gu, Yanzhao and Law, KL Eddie and Cheong, Ngai},
  booktitle={2024 22nd International Symposium on Modeling and Optimization in Mobile, Ad Hoc, and Wireless Networks (WiOpt)},
  pages={124--131},
  year={2024},
  organization={IEEE}
}

@article{gwoa,
  title={GWOA: A multi-strategy enhanced whale optimization algorithm for engineering design optimization},
  author={Gu, Yanzhao and Wei, Junhao and Li, Zikun and Lu, Baili and Pan, Shirou and Cheong, Ngai},
  journal={Plos one},
  volume={20},
  number={9},
  pages={e0322494},
  year={2025},
  publisher={Public Library of Science San Francisco, CA USA}
}

@article{drrt,
  title={Research on UAV Applications in Public Administration: Based on an Improved RRT Algorithm},
  author={Xie, Zhanxi and Lu, Baili and Gu, Yanzhao and Li, Zikun and Wei, Junhao and Cheong, Ngai},
  journal={arXiv preprint arXiv:2508.14096},
  year={2025}
}

@article{lsewoa,
  title={LSEWOA: An Enhanced Whale Optimization Algorithm with Multi-Strategy for Numerical and Engineering Design Optimization Problems},
  author={Wei, Junhao and Gu, Yanzhao and Yan, Yuzheng and Li, Zikun and Lu, Baili and Pan, Shirou and Cheong, Ngai},
  journal={Sensors},
  volume={25},
  number={7},
  pages={2054},
  year={2025},
  publisher={MDPI}
}

@article{mrbmo,
  title={MRBMO: An Enhanced Red-Billed Blue Magpie Optimization Algorithm for Solving Numerical Optimization Challenges},
  author={Lu, Baili and Xie, Zhanxi and Wei, Junhao and Gu, Yanzhao and Yan, Yuzheng and Li, Zikun and Pan, Shirou and Cheong, Ngai and Chen, Ying and Zhou, Ruishen},
  journal={Symmetry},
  volume={17},
  number={8},
  pages={1295},
  year={2025},
  publisher={Multidisciplinary Digital Publishing Institute}
}

@inproceedings{tswoa,
  title={TSWOA: An Enhanced WOA with Triangular Walk and Spiral Flight for Engineering Design Optimization},
  author={Wei, Junhao and Gu, Yanzhao and Yan, Yuzheng and Wang, Yapeng and Li, Zikun and Lu, Baili and Pan, Shirou and Cheong, Ngai},
  booktitle={2025 8th International Conference on Advanced Algorithms and Control Engineering (ICAACE)},
  pages={186--194},
  year={2025},
  organization={IEEE}
}

@article{lswoa,
  title={LSWOA: An enhanced whale optimization algorithm with Levy flight and Spiral flight for numerical and engineering design optimization problems},
  author={Wei, Junhao and Gu, Yanzhao and Xie, Zhanxi and Yan, Yuzheng and Lu, Baili and Li, Zikun and Cheong, Ngai and Zhang, Jiafeng and Zhang, Song},
  journal={Plos one},
  volume={20},
  number={9},
  pages={e0322058},
  year={2025},
  publisher={Public Library of Science San Francisco, CA USA}
}

@article{rwoa,
  title={RWOA: A novel enhanced whale optimization algorithm with multi-strategy for numerical optimization and engineering design problems},
  author={Wei, Junhao and Gu, Yanzhao and Lu, Baili and Cheong, Ngai},
  journal={PloS one},
  volume={20},
  number={4},
  pages={e0320913},
  year={2025},
  publisher={Public Library of Science San Francisco, CA USA}
}

@article{ahrrt,
	doi = {10.20944/preprints202511.1805.v1},
	url = {https://doi.org/10.20944/preprints202511.1805.v1},
	year = 2025,
	month = {November},
	publisher = {Preprints},
	author = {Junhao Wei and Yanzhao Gu and Ran Zhang and Wenxuan Zhu and Shuai Wu and Yapeng Wang and Ngai Cheong and Zhiwen Wang and Sio-Kei Im and Xu Yang},
	title = {AHRRT: An Enhanced Rapidly-Exploring Random Tree Algorithm with Heuristic Search for UAV Urban Path Planning},
	journal = {Preprints}
}

@inproceedings{xgb,
  title={Credit Card Fraud Detection Based on MiniKM-SVMSMOTE-XGBoost Model},
  author={Gu, Yanzhao and Wei, Junhao and Cheong, Ngai},
  booktitle={Proceedings of the 2024 8th International Conference on Big Data and Internet of Things},
  pages={252--258},
  year={2024}
}

@article{bib1,
  title={Exploring the Development of Personalizing Learning in Chinese Higher Education: A Systematic Review of Cognitive Evolution Engine by AI},
  author={Huang, Mingjing and Cheong, Ngai and Zhang, Zhuofan and Liu, Jiaqi},
  journal={IEEE Transactions on Learning Technologies},
  year={2025},
  publisher={IEEE}
}

@article{xgbwoa,
  title={Performance evaluation of hybrid WOA-XGBoost, GWO-XGBoost and BO-XGBoost models to predict blast-induced ground vibration},
  author={Qiu, Yingui and Zhou, Jian and Khandelwal, Manoj and Yang, Haitao and Yang, Peixi and Li, Chuanqi},
  journal={Engineering with Computers},
  volume={38},
  number={Suppl 5},
  pages={4145--4162},
  year={2022},
  publisher={Springer}
}

@article{IWOA,
  title={IWOA-RNN: An improved whale optimization algorithm with recurrent neural networks for traffic flow prediction},
  author={Liu, Zhiyou and Li, Xinbin and Lu, Zhigang and Meng, Xianhui},
  journal={Alexandria Engineering Journal},
  volume={117},
  pages={563--576},
  year={2025},
  publisher={Elsevier}
}

@article{glnwoa,
  title={An Enhanced Whale Optimization Algorithm with Log-Normal Distribution for Optimizing Coverage of Wireless Sensor Networks},
  author={Wei, Junhao and Gu, Yanzhao and Zhang, Ran and Li, Yanxiao and Zhu, Wenxuan and Wang, Yapeng and Yang, Xu and Cheong, Ngai},
  journal={arXiv preprint arXiv:2511.15970},
  year={2025}
}

\end{document}